\documentclass[12pt]{article}
\usepackage{epsf}
\begin{document}
\def\be{\begin{equation}}
\def\ee{\end{equation}}
\def\ba{\begin{array}{l}}
\def\ea{\end{array}}
\def\bea{\begin{eqnarray}}
\def\eea{\end{eqnarray}}
\def\eq#1{(\ref{#1})}
\def\fig#1{Fig \ref{#1}} 
\def\wgnc{\bar{\wedge}}
\def\del{\partial}
\def\der{\overline \del}
\def\wg{\wedge}
\def\bull{$\bullet$}
\def\gap{\vspace{10ex}}
\def\tgap{\vspace{3ex}}
\def\sgap{\vspace{5ex}}
\def\lgap{\vspace{20ex}}
\def\half{\frac{1}{2}}
\def\pto{\vfill\eject}
\def\gst{g_{\rm st}}
\def\tC{{\widetilde C}}
\def\z{{\bar z}}
\def\o{{\cal O}}
\def\J{{\cal J}}
\def\S{{\cal S}}
\def\X{{\cal X}}
\def\N{{\cal N}}
\def\A{{\cal A}}
\def\H{{\cal H}}
\def\D{{\tilde D}}
\def\d{{\cal D}}
\def\re#1{{\bf #1}}
\def\nn{\nonumber}
\def\nl{\hfill\break}
\def\ni{\noindent}
\def\bibi{\bibitem}
\def\c#1{{\hat{#1}}}
\def\eps{{\epsilon}}
\pretolerance=1000000
\begin{flushright}
TIFR/TH/00-44\\
August 2000\\
\end{flushright}
\begin{center}
\vspace{2 ex}
{\large\bf A Note on Gauge Invariant Operators in Noncommutative
Gauge Theories and the Matrix Model}
\\
\vspace{3 ex}
Avinash Dhar and Spenta R. Wadia \\
~\\
{\sl Department of Theoretical Physics,}\\
{\sl Tata Institute of Fundamental Research, }\\
{\sl Homi Bhabha Road, Mumbai 400 005, INDIA.} \\
\vspace{1 ex}
{\sl adhar@tifr.res.in,wadia@tifr.res.in}

\vspace{10 ex}
\pretolerance=1000000
\bf ABSTRACT\\
\end{center}
\vspace{1 ex} 

In this note we discuss local gauge-invariant operators in
noncommutative gauge theories. Inspired by the connection of these
theories with the Matrix model, we give a simple construction of a
complete set of gauge-invariant operators. We make connection with the
recent discussions of candidate operators which are dual to closed
strings modes. We also discuss large Wilson loops which in the limit
of vanishing noncommutativity, reduce to the closed Wilson loops of 
the ordinary gauge theory.

\vfill

\clearpage

\vspace{8ex}

\section{\large Introduction} 

One question of intrinsic interest in any gauge theory is the construction of
local gauge-invariant operators since these form a complete set of observables
of the theory.  This question has been addressed in several recent works
\cite{witten2,malda-russo,hash-itz,dmwy,trivedi,AIIKKT,IIKK,szabo,gopa,
gross-nekra,gaume-wad} in the context of noncommutative gauge theories.  In
these theories the construction of gauge-invariant operators is made somewhat
nontrivial by the fact that simple analogues of local gauge-invariant operators
of commutative gauge theories, taken over to the noncommutative case, are not
local, but are integrated over all of space.  It turns out, however, that more
general gauge-invariant operators do exist and were constructed in \cite{IIKK}.
The set of gauge-invariant operators in noncommutative gauge theories
constructed in this work has been further discussed in
\cite{ReyU,DasR,gross-hash-itz}.  Furthermore, operators presented in
\cite{gross-hash-itz} have the desirable property that they reproduce the
corresponding local operators in ordinary commutative gauge theories at scales
large compared to the noncommutativity scale.

In this note we exploit the connection
\cite{BSS,li,fato,schia,cornalba,ishi,IIKK,oleg,bars-minic,szabo} of the matrix
model to noncommutative gauge theories (emphasized recently by Seiberg
\cite{sei}) to present a particularly simple construction of local
gauge-invariant operators in these theories.  The matrix model has an underlying
$U(N)$ gauge symmetry and it is easy to write operators invariant under this
symmetry.  Dp-branes arise in the matrix model as classical solutions in the
limit of large $N$ and the $U(N)$ gauge symmetry of the matrix model reappears
on the branes as the noncommutative gauge symmetry.  Because of this connection
$U(N)$-invariant operators of the matrix model reappear as gauge-invariant
operators on the branes.  This fact directly leads to a simple construction of
local gauge-invariant operators in noncommutative gauge theories.

\section{\large Dp-branes in Matrix Model} 

The dynamical variables in the matrix model \cite{BFSS} are nine $N\times N$
hermitian matrices $X^I$, $I=1,2,...,9$. Time-independent bosonic
classical solutions are determined by extremizing the potential
Tr$[X^I, X^J]^2$, which gives the equation of motion 
\bea 
\label{twoone}
[X_J, [X^I, X^J]]=0.
\eea
In the large N limit Dp-branes correspond to the solutions 
of this equation given by \cite{BSS}
\bea
\label{twotwo}
X^i &=& x^i, \ i=1,2,...,p, \nn \\
X^I &=& 0, \ I=p+1,...,9.
\eea
where
\bea
\label{twothree}
[x^i, x^j]=i\theta^{ij},
\eea
The rank of the matrix $\theta$ is $p$ for Dp-branes so that one has
maximal noncommutativity. Also note that in the matrix model $p$ is
even.

Expanding $X^I$ around the above classical solution gives a
noncommutative gauge theory on the branes. If the classical solution
represents $Q$ Dp-branes, one gets a $U(Q)$ noncommutative gauge
theory. 

Let us parameterize the fluctuations as 
\bea
\label{twofour}
X^i = x^i + \theta^{ij} \hat A_j(x^i)
\eea
We have written the fluctuations as dependent on $x^i$ because we can 
expand it in terms of a complete set of operators. An example, in the large N limit, is the Weyl basis defined by the exponential operators
$g(\alpha)=\exp{i\alpha_ix^i}$.
We also have the scalar fluctuations $\phi^a = X^{a+p}, \
a=1,2,...,9-p$. 

Now, the time independent $U(N)$ gauge symmetry of the matrix model
\footnote{In the IKKT matrix model \cite{IKKT} this will be the full gauge symmetry} 
acts on $X^I$ as $X^I \rightarrow U X^I U^\dagger$. This action
descends on the fluctuations $\hat A_i$ on the branes as $U(Q)$
noncommutative gauge symmetry
\bea
\hat A_i \rightarrow U \hat A_i U^\dagger + iU\der_iU^\dagger
\label{ncgt}
\eea
where 
\be
\der_i=-i\theta^{-1}_{ij}adx_j
\label{der}
\ee
provided we transfer, as above, the $U(N)$ transformation of the 
background to the fluctuation. Here $\theta^{-1}$ is the inverse of the matrix $\theta$. 

The transformation of the fluctuation in (\ref{ncgt}) is indeed a
gauge transformation because the second term can be understood as a
parallel transport \cite{gaume-wad, gross-nekra} defined in terms of
the derivative operator of the non-commutative gauge theory.  From
here it is clear that the gauge group of the non-commutative gauge
theory is inherited from the large N limit of the matrix model, and
also the definition (\ref{der}) makes it clear that translations are
generated by $U(\infty)$ rotations. Thus, we see that the set of
gauge-invariant operators on the branes is identical to and directly
inherited from the gauge-invariant operators in the matrix model. Also
gauge invariant operators are necessarily translation invariant.

\section{\large Gauge Invariant Operators}

Gauge-invariant operators are very easy to construct in the matrix
model because its $U(N)$ gauge symmetry is linearly realized on the
dynamical variables $X^I$. Thus, the matrix trace of an arbitrary
function of $X^I$ is gauge-invariant. A complete set of
gauge-invariant operators is given by ${\rm Tr}(X^{I_1}...X^{I_n})$
where $n \leq N$, since for $n \geq N$ the trace can be rewritten in
terms of traces of fewer number of matrices. In the large $N$ limit,
which is what we need in the Dp-brane classical background, it is more
convenient to consider the operators
\bea
\hat O_k = e^{ik.X}.
\label{threeone}
\eea 
Traces of products of these operators for different values of $k$ can
be used to generate all the above gauge-invariant operators for finite
$N$. For example, $${\rm Tr} [X^I,X^J]^2 = -2(\del_I {\del'}_J -\del_J
{\del'}_I)^2 \ {\rm Tr}(e^{ik.X} e^{ik'.X})|_{k=k'=0},$$ where $\del$
and $\del'$ are derivatives with respect to $k$ and $k'$. That is, all
the gauge-invariant operators of the matrix model are contained in the
set of operators
\bea
\hat O_{kk'...} = e^{ik.X} \ e^{ik'.X} \cdots
\label{threetwo}
\eea 
By the correspondence of the matrix model $U(N)$ gauge symmetry with
the noncommutative gauge symmetry on the branes discussed above, then,
this set of operators must reproduce all the gauge-invariant operators
in the noncommutative gauge theory on the branes. In the rest of this
note we will show that this is indeed the case.

\subsection{Straight and Curved Wilson lines}

Traces of operators of the type in (\ref{threeone}) can be given the
interpretation of straight Wilson lines in the noncommutative gauge
theory on the branes. To see how that comes about, let us write 
\bea
\hat O_k &=& e^{ik.X} \nn \\
&=& e^{i{k \over n}.X} \ e^{i{k \over n}.X} \cdots 
(n \ {\rm factors}).
\label{threethree}
\eea 
At the end we will take the limit $n \rightarrow \infty$. 
Here we will restrict ourselves to $k_I$ with non-zero components only in the brane directions. The more general case is treated in the next sub-section.
Let us now
use (\ref{twofour}) to write, in an obvious notation, 
\bea
e^{i{k \over n}.X} &=& e^{i\epsilon\theta^{-1}.x + i \epsilon.\hat A(x)} 
\nn \\ 
&=&e^{i\epsilon\theta^{-1}.x} \ e^{i\epsilon.\hat A(x) + o(\epsilon^2)}
\nn
\eea
where $\epsilon = {k \over n}\theta$. Using this in
(\ref{threethree}), we get
\bea
\hat O_k &=& e^{i\epsilon\theta^{-1}.x} \ e^{i\epsilon.\hat A(x) + 
o(\epsilon^2)} \ e^{i\epsilon\theta^{-1}.x} \ e^{i\epsilon.\hat A(x) + 
o(\epsilon^2)} \cdots (n \ {\rm factors}) \nn \\
&=& e^{i\epsilon.\hat A(x + \epsilon) + o(\epsilon^2)} \
e^{i\epsilon.\hat A(x + 2 \epsilon) + o(\epsilon^2)} \
\cdots e^{i\epsilon.\hat A(x + n \epsilon) + o(\epsilon^2)} \
e^{i n \epsilon\theta^{-1}.x}
\label{threefour}
\eea

In arriving at the last line above we have used the fact that the
adjoint action of $x$ generates a translation by virtue of the algebra
(\ref{twothree}). In the limit $n \rightarrow \infty$, the product of
operators involving the gauge field in the last line gives the Wilson
line operator $U(x, x + k \theta)$
along a straight line path given by the vector $k \theta$. Thus the
gauge-invariant operator Tr$\hat O_k$ of the matrix theory translates
into the Wilson line Tr$U(x, x + k \theta) \ e^{ik.x}$ in the
noncommutative gauge theory on the branes. 

Upto a phase, the more general operators $\hat O_{kk'...}$ given in
(\ref{threetwo}) can similarly be interpreted in the noncommutative
gauge theory on the branes as Wilson line operators along a general
curved path determined by the straight line segments given by the
vectors $k \theta$, $k' \theta$, etc. That is, 
\bea 
\hat O_{kk'...} = U(x, x + k \theta) \ U(x + k \theta, x + (k + k') \theta) 
\cdots e^{ik.x} \ e^{ik'.x} \cdots
\label{threefive}
\eea

\vfill
\newpage

An example of such a Wilson line has been shown in Fig. 1. Since any
continuous curve can be approximated by straight line segments one can
easily construct the Wilson line corresponding to an arbitrary open
curve.

\vspace{5cm}
\includegraphics{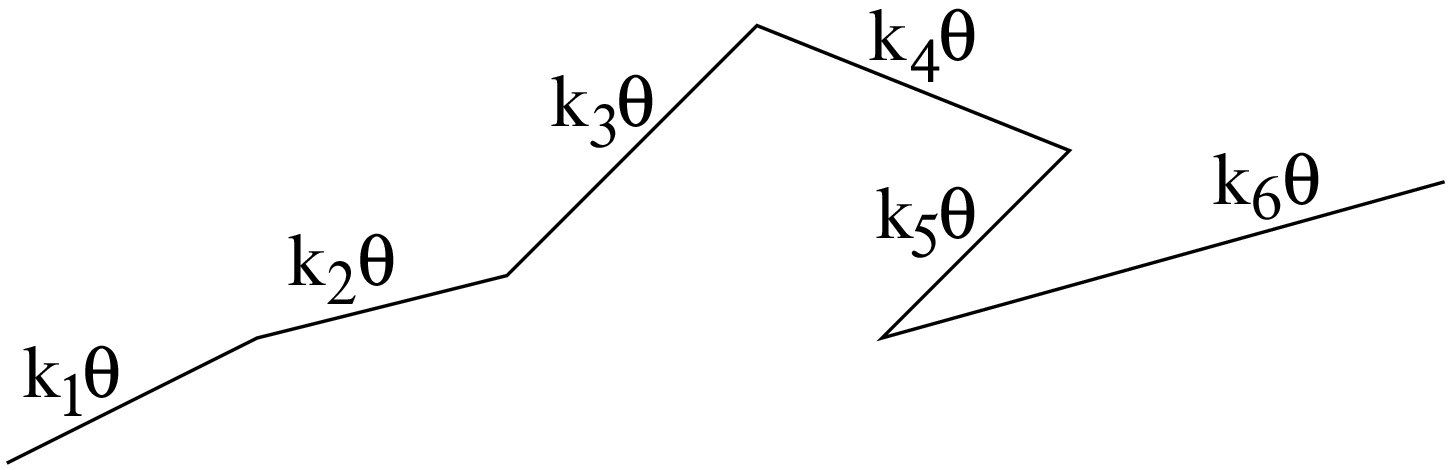} 
\begin{center}
Fig. 1: Polygonal Wilson line
\end{center}

Note that if $x_i$ is a brane background then so is $x_i+a_iI$ where $I$ is the
identity operator and $a_i$ is a real number.  This fact immediately implies
that the expectation value of $Tr \hat O_{kk'...}$ conserves momentum and is
non-zero only if the sum $k + k'+ ....  = 0$.

\subsection{Generalised Wilson Lines}

Let us now consider the more general situation of arbitrary $k_I$,
i.e. $k_I = (k_i, s_a)$ where $i=1,2,...,p$ and $a=1,2,...,9-p$. In this case proceeding after eqn. (\ref{threethree}), we have 
\be
e^{i{k \over n}.X} = e^{i\epsilon\theta^{-1}.x + i \epsilon^i\hat A(x)_i
\nn \\
+ i \frac{s_a}{n}\hat \phi_{a}(x)} 
\label{threethree'}
\ee
where as before 
$\epsilon^i = {k_j\over n}\theta^{ji}$.
Thus the analogue of (\ref{threefour}) is 
\bea
\hat O_k &=& e^{i\epsilon^i\hat A_i(x + \epsilon) + 
i \frac{s_a}{n} \hat \phi_{a}(x + \epsilon) + o({1\over n^2})} \
e^{i\epsilon^i\hat A_i(x + 2\epsilon) + 
i \frac{s_a}{n}\hat \phi_{a}(x + 2\epsilon) + o({1\over n^2})} \nn \\
&&\cdots e^{i\epsilon^i\hat A_i(x + n \epsilon) + i \frac{s_a}{n} \hat \phi_{a}(x + n\epsilon) + o({1\over n^2})} \
e^{i n \epsilon\theta^{-1}.x}
\label{threefour'}
\eea
In the limit $n\rightarrow \infty$ this gives rise to a modified
Wilson line operator $U_{s}(x, x + k \theta)$ along the straight line
path given by the vector $k\theta$. The path is now  
characterized by the additional `internal' quantities $s_a, s_a',....$. These
generalized Wilson line operators are similar to those introduced by
Maldacena \cite{maldaloop} in the context of AdS/CFT. It would be nice
to understand the connection more quantitatively.

\subsection{Operators Dual to the Closed String Modes}

When the sum of momenta $(k + k' + \cdots)$ vanishes, the operator
$\hat O_{kk'...}$ corresponds, on the branes, to a closed Wilson loop
(untraced), which was defined e.g. in \cite{gaume-wad} with 
$\epsilon^m_i = (k^m\theta)_i$ being the length of the mth loop segment.
One can let $m \rightarrow \infty $
and $k^m\rightarrow 0$ for fixed $\theta$ 
in such a way that the polygonal loop becomes a
continuous curve with no net momentum. 
Of course, in general one also includes here operators that traverse the
geometrical loop several times.

Now, a special class of operators arises when an operator, which corresponds to
one or more closed Wilson loops, is present anywhere in the product of
exponentials that defines the general operator $\hat O_{kk'...}$.  On the
branes, such an operator corresponds to a Wilson line with a closed polygon loop
somewhere along it.  An example of such a Wilson line with a single closed loop
is shown in Fig 2.  In the case of a single closed loop, the position of the
loop along the Wilson line does not matter.  This is because we are dealing with
the trace of the operator (\ref{threetwo}) in which a certain sequence of
momenta adds up to zero.

\vspace{5cm}
\includegraphics{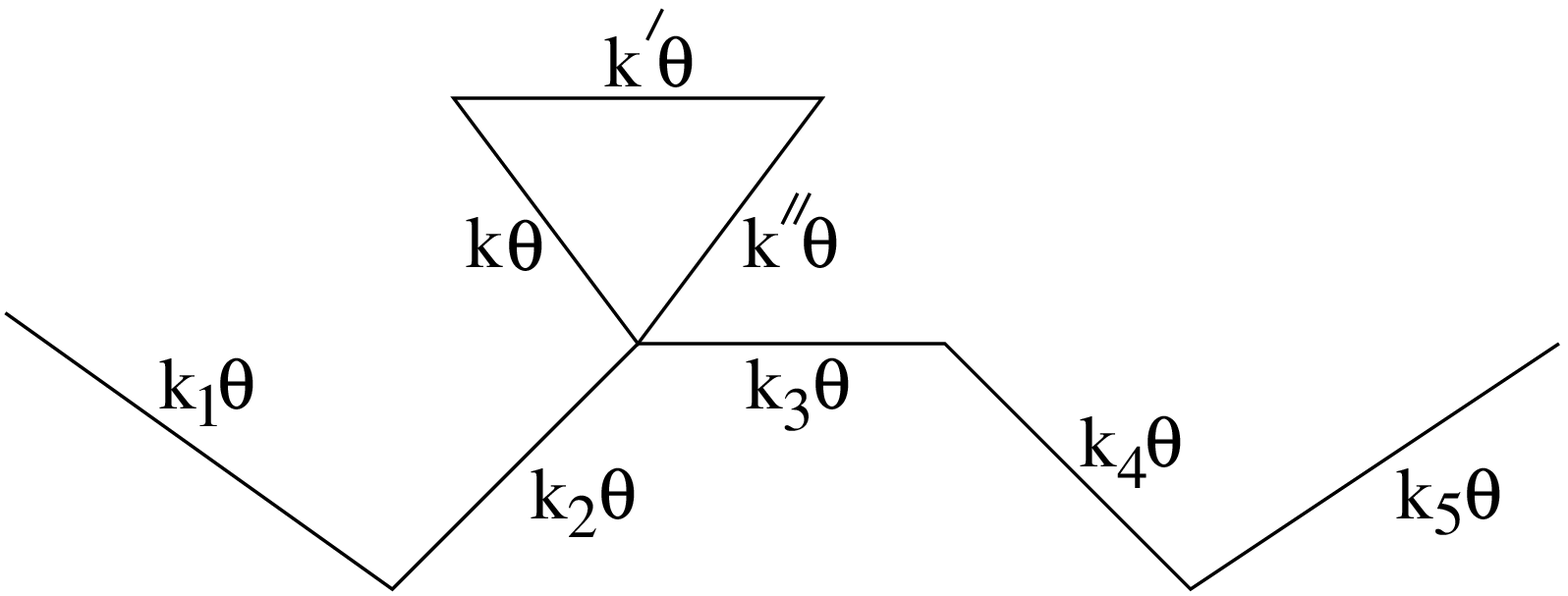} 
\begin{center}
Fig. 2: Wilson line with an attached closed loop
\end{center}

The above constructions are easily generalized to continuous curves.  If the
open curve is a straight line specified by a fixed momentum and the closed loop
is vanishingly small, the Wilson line generates the gauge-invariant operators
that have recently been discussed in \cite{gross-hash-itz} as candidate
operators dual to supergravity modes.  As an example, consider the following
operator
\bea
\hat O_{k_1kk'k''k_2} =  e^{ik_1.X} \ e^{ik.X} \ e^{ik'.X} \ e^{ik''.X} 
\ e^{ik_2.X}
\label{threesix}
\eea
where $(k + k' + k'')=0$ and $k$ and $k'$ are small, though $k_1$ and
$k_2$ need not be small. Writing $k''=-(k+k')$ we have
\bea
e^{ik.X} \ e^{ik'.X} \ e^{-i(k + k').X} = 1 - {1 \over 2} [k.X, k'.X] + 
O(k^3).
\label{threeseven}
\eea
Thus we get
\bea
\hat O_{k_1kk'k''k_2} = e^{ik_1.X} \ e^{ik_2.X} - {1 \over 2} k_I k'_J \
e^{ik_1.X} [X^I,X^J] e^{ik_2.X} + O(k^3).
\label{threeeight}
\eea 

It is also instructive to note that if $(k + k' + k'')$ were non-zero
in $\hat O_{k_1kk'k''k_2}$ (open loop), then we could have defined
$\hat O_{k_1kk'k''pqk_2}$ such that $p+q=0$ and $k+k'+k''+p=0$. This
is just the same as $\hat O_{k_1kk'k''k_2}$ but now can be interpreted
as a closed loop placed along an open line. Taking the trace would
once more move the loop anywhere along the open line. This simple
example is illustrative of the fact that in many cases of
interest (though not all) open Wilson lines are equivalent to straight
Wilson lines with loops attached.

If the loop is not small, Wilson lines of the above type with fixed momentum
(Fig. 3) would then seem to correspond to the {\sl full closed string} in the
dual theory.  This is the analogue of closed Wilson loop in ordinary gauge
theory and reduces to it in the limit of vanishing noncommutativity ($\theta
\rightarrow 0$).  It would be interesting to compute the expectation value of
the Wilson loop in the large $\theta$ limit using Maldacena duality, much in the
same spirit the calculation is done in the absence of the B-field.  Also since
Wilson loops create electric flux lines, it would be of interest to study the
connection with the work of \cite{harvey} in the case of the $D_3$ branes.

\vspace{5cm}
\includegraphics{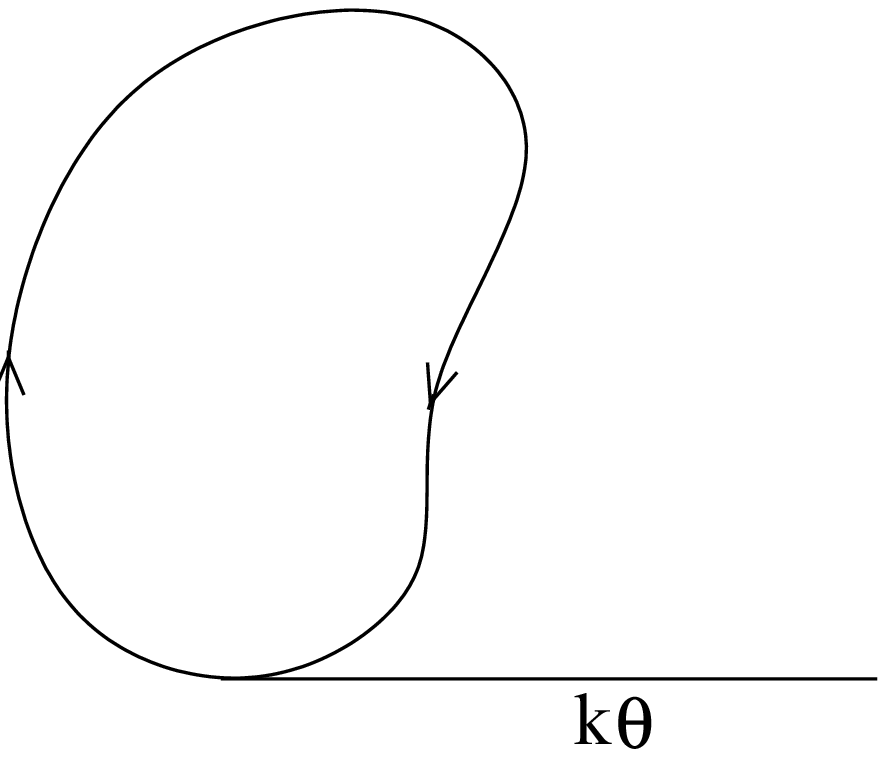} 
\begin{center}
Fig. 3: Large Wilson loop attached to a Wilson line of momentum $k$
\end{center}


\begin{thebibliography}{999}

\bibitem{witten2} 
N.~Seiberg and E.~Witten,  
``Non-commutative Geometry and String theory'', 
JHEP ~9909:032 ~(1999), hep-th/9908142. 
\bibitem{malda-russo}
J.~Maldacena and J.~Russo
"Large N Limit of Non-Commutative Gauge Theories",
hep-th/9908134.
\bibitem{hash-itz}
A.~Hashimoto, N.~Itzhaki,
"Non-Commutative Yang-Mills and the AdS/CFT Correspondence",
hep-th/9907166. 
\bibitem{dmwy}   
A.~Dhar, G.~Mandal, S.~R.~Wadia and K.~P.~Yogendran, 
``D1/D5 system with B-field, noncommutative geometry and the CFT of the  Higgs branch,'' 
hep-th/9910194. 
\bibitem{trivedi}
S.~Das, S.~Kalyana Rama, S.~Trivedi,
"Supergravity with Self-dual B fields and Instantons in Noncommutative Gauge Theory",
hep-th/9911137.
\bibitem{AIIKKT}
H.~Aoki, N.~Ishibashi, S.~Iso, H.~Kawai, Y.~Kitazawa, T.~Tada,
``Noncommutative Yang-Mills in IIB Matrix Model'', 
Nucl.Phys. {\bf B565} 176 (2000) , hep-th/9908141.
\bibitem{IIKK}
N.~Ishibashi, S.~Iso, H.~Kawai, Y.~Kitazawa,
``Wilson Loops in Noncommutative Yang Mills'',
Nucl.Phys. {\bf B573} 573 (2000), hep-th/9910004.
\bibitem{szabo}
J.~Ambjorn, Y.M.~Makeenko, J.~Nishimura, R.J.~Szabo,
``Finite N Matrix Models of Noncommutative Gauge Theory'',
JHEP 9911 (1999) 029, hep-th/9911041; ``Nonperturbative Dynamics of Noncommutative Gauge Theory'', Phys.Lett. {\bf B480} 399 (2000), hep-th/0002158;
J.~Ambjorn, Y.M.~Makeenko, J.~Nishimura, R.J.~Szabo,
``Lattice Gauge Fields and Discrete Noncommutative Yang-Mills Theory'',
hep-th/0004147.
\bibitem{gopa}
R.~Gopakumar, S.~Minwalla, A.~Strominger, 
''Noncommutative Solitons'', hep-th/0003160. 
\bibitem{gross-nekra}
D.~J.~Gross and N.~Nekrasov,
``Monopoles and Strings in Non-Commutative Gauge Theory'',
hep-th/0005204.
\bibitem{gaume-wad}
L.~Alvarez-gaume and S.~R.~Wadia,
``Gauge Theory on a Quantum Phase Space'',
hep-th/0006219.
\bibitem{maldaloop}
J.~Maldacena,
``Wilson loops in large N field theories'',
Phys.Rev.Lett. {\bf 80 } 4859 (1998), hep-th/9803002.
\bibitem{ReyU}
Soo-Jong Rey and R. von Unge,
"S-Duality, Noncritical Open String and Noncommutative Gauge Theory",
hep-th/0007089.
\bibitem{DasR}
S.~Das and Soo-Jong Rey,
"Open Wilson Lines in Noncommutative Gauge Theory and Tomography of Holographic Dual Supergravity",
hep-th/0008042.
\bibitem{gross-hash-itz}
D.~J. Gross, A.~Hashimoto, N.~Itzhaki,
"Observables of Non-Commutative Gauge Theories",
hep-th/0008075.
\bibitem{BSS}
T.~Banks, N.~Seiberg and S.~Shenker,
"Branes from Matrices",
Nucl. Phys. {\bf B490} 91 (1997), hep-th/9612157.
\bibitem{li}
M.~Li,
"Strings from Type II matrices",
Nucl. Phys. {\bf B499} 149 (1997),
hep-th/9612222.
\bibitem{fato}
A.~Fatollahi,
``Gauge Symmetry As Symmetry Of Matrix Coordinates'',
hep-th/0007023.
\bibitem{schia}
L.~Cornalba and R.~Schiappa, 
``Matrix Theory Star Products from the Born-Infeld Action'',
hep-th/9907211. 
\bibitem{cornalba}
L.~Cornalba,
``D-brane Physics and Noncommutative Yang-Mills Theory'',hep-th/9909081.
\bibitem{ishi}
 N.~Ishibashi,
``A Relation between Commutative and Noncommutative Descriptions of D-branes'', hep-th/9909176.
\bibitem{bars-minic}
I.~Bars and N.~Minic,
"Non-commutative Geometry on a Discrete Periodic Lattice
and Gauge Theory",
hep-th/9910091.
\bibitem{oleg}
O.~Andreev and H.~Dorn,
``On Open String Sigma-Model and Noncommutative Gauge Fields'',
hep-th/9912070.
\bibitem{sei}
N.~Seiberg,
"A Note on Background Independence in Noncommutative Gauge Theories, 
Matrix Model and Tachyon Condensation", hep-th/0008013.
\bibitem{BFSS} 
T.~Banks, W.~Fischler, S.H.~Shenker, L.~Susskind 
``M-theory as a Matrix model: a conjecture.'' 
Phys.~Rev.~{\bf D55} 5112 (1997), hep-th/9610043. 
\bibitem{IKKT}
N.~Ishibashi, H.~Kawai, Y.~Kitazawa, A.~Tsuchiya
"A Large-N Reduced Model as Superstring",
hep-th/9612115.
\bibitem{harvey} 
J.~Harvey, P.~Kraus, F.~Larsen and E.~Martinec, 
``D-branes and Strings as Non-commutative Solitons'', hep-th/0005031. 
\end{thebibliography}
\end{document}